\documentclass[pdflatex,sn-nature]{sn-jnl}

\usepackage{graphicx}%
\usepackage{multirow}%
\usepackage{amsmath,amssymb,amsfonts}%
\usepackage{amsthm}%
\usepackage{mathrsfs}%
\usepackage[title]{appendix}%
\usepackage{xcolor}%
\usepackage{textcomp}%
\usepackage{manyfoot}%
\usepackage{booktabs}%
\usepackage{algorithm}%
\usepackage{algorithmicx}%
\usepackage{algpseudocode}%
\usepackage{listings}%
\usepackage{upgreek}

\theoremstyle{thmstyleone}%
\theoremstyle{thmstyletwo}%

\theoremstyle{thmstylethree}%

\begin{document}

\title[Article Title]{Integrated Spectropolarimeter by Metasurface-Based Diffractive Optical Networks}


\author[1,2]{\fnm{Jumin} \sur{Qiu}}

\author[3,4]{\fnm{Tingting} \sur{Liu}}

\author[3]{\fnm{Chenxuan} \sur{Xiang}}

\author*[1,2]{\fnm{Tianbao} \sur{Yu}}\email{yutianbao@ncu.edu.cn}

\author*[3]{\fnm{Qiegen} \sur{Liu}}\email{liuqiegen@ncu.edu.cn}

\author*[3,4]{\fnm{Shuyuan} \sur{Xiao}}\email{syxiao@ncu.edu.cn}

\affil[1]{\orgdiv{School of Physics and Materials Science}, \orgname{Nanchang University}, \orgaddress{\city{Nanchang}, \postcode{330031}, \state{Jiangxi}, \country{China}}}

\affil[2]{\orgdiv{Jiangxi Provincial Key Laboratory of Photodetectors}, \orgname{Nanchang University}, \orgaddress{\city{Nanchang}, \postcode{330031}, \state{Jiangxi}, \country{China}}}

\affil[3]{\orgdiv{School of Information Engineering}, \orgname{Nanchang University}, \orgaddress{\city{Nanchang}, \postcode{330031}, \state{Jiangxi}, \country{China}}}

\affil[4]{\orgdiv{Institute for Advanced Study}, \orgname{Nanchang University}, \orgaddress{\city{Nanchang}, \postcode{330031}, \state{Jiangxi}, \country{China}}}


\abstract{Conventional spectrometer and polarimeter systems rely on bulky optics, fundamentally limiting compact integration and hindering multi-dimensional optical sensing capabilities. Here, we propose a spectropolarimeter enabled by metasurface-based diffractive optical networks that simultaneously performs spectrometric and polarimetric measurements in a compact device. By leveraging the wavelength- and polarization-dependent phase modulation of metasurfaces, our system encodes the spectral and polarization information of incident light into spatially resolved intensity distributions, which are subsequently decoded by a trained deep neural network, enabling simultaneous high-accuracy reconstruction of both spectral compositions and Stokes parameters through a single-shot measurement. Experiments validate the proposed network’s accurate reconstruction of the spectral and polarization information across a broad wavelength range, and further confirm its imaging capability. Notably, we demonstrate a chip-integrated sensor prototype combing both measurement functionalities into a commercial CMOS image sensor. This integrated platform provides a compact solution for on-chip multi-dimensional optical sensing, holding significant potential for versatile sensing, biomedical diagnosis, and industrial metrology.}

\maketitle

\section{Introduction}\label{sec1}
Optical sensing technologies, particularly those employing spectrometers and polarimeters, play a central role in modern scientific and industrial applications, ranging from remote sensing and biomedical diagnostics to materials characterization and machine vision\cite{Yuan2023, He2021}. Spectrometers extract critical chemical and physical properties of analytes through spectral signatures, while polarimeters probe structural and optical characteristics (e.g., birefringence, dichroism, surface morphology) by quantifying polarization states via Stokes parameters. However, conventional approaches typically require bulky discrete devices or sequential measurement steps to acquire spectral and polarization information. This fundamental limitation hinders the rapid acquisition of comprehensive optical properties in dynamic scenarios and impedes the development of compact, integrated sensing platforms\cite{Yang2021, Li2022, Xue2024}.

Recent advances in metasurface technology offer promising pathways toward miniaturized optical sensing\cite{Dorrah2022}. These subwavelength nanostructures can replace conventional bulky optics, enabling high-density integration and unprecedented wavefront control through tailored amplitude\cite{Zheng2021, Deng2024, Liu2024}, phase\cite{Deng2018, Li2024}, polarization\cite{Ding2020, Wang2023, Li2025a}, as well as orbital angular momentum\cite{Ren2020, Meng2025}. For spectrometry, notable examples include spatial-spectral encoding\cite{Wang2023a, Cai2024, Zhang2025, Lee2025}, metalenses for snapshot hyperspectral imaging\cite{Lin2023, Kim2023, Audhkhasi2025}, and quasi-bound states in the continuum enabling miniaturized spectrometers\cite{Tittl2018, Tang2024}. 
Similarly, for polarimetry, metasurface-based solutions have demonstrated remarkable capabilities, such as polarization-multiplexed focusing and holograms\cite{Li2020, Liu2022a, Ou2022, Wang2024}, Stokes parameter mapping to intensity patterns\cite{Hu2024, Chen2023a, Yang2025}, and metasurfaces for snapshot polarimetric imaging\cite{Zuo2023, Shen2023, Zaidi2024, He2024, Li2025}.
Despite significant progress, existing metasurface devices remain functionally specialized, primarily performing either spectrometric or polarimetric measurement. Integrating both functionalities presents inherent challenges, often necessitating complex optical encoding schemes, multi-step measurements, or intricate reconstruction algorithms\cite{Ni2022, Chen2024, Zhang2024}. This complexity can complicate practical implementation and limit adaptability.

To address these challenges, we propose an integrated spectropolarimeter by exploiting the concept of diffractive optical networks (DONs), which contains cascaded metasurfaces. 
Our system utilizes the jointly optimized, wavelength- and polarization-dependent phase modulation of metasurfaces to encode the spectral and polarization information of incident light into a single spatially resolved intensity pattern. A trained deep neural network subsequently decodes this pattern, enabling the simultaneous reconstruction of spectral compositions and full Stokes parameters from a single-shot measurement. Experimental results validate high-accuracy spectral resolution and polarization state reconstruction across a broad wavelength range, and further confirm imaging capability. Finally, we implement a chip-integrated sensor prototype, translating the metasurface DONs onto a commercial CMOS image sensor. This integration eliminates moving parts and discrete optical components, significantly reducing system complexity while enhancing robustness. 

\section{Results}\label{sec2}

\subsection{Spectropolarimeter system design}

\begin{figure*}[!ht]%
\centering
\includegraphics[width=\textwidth]{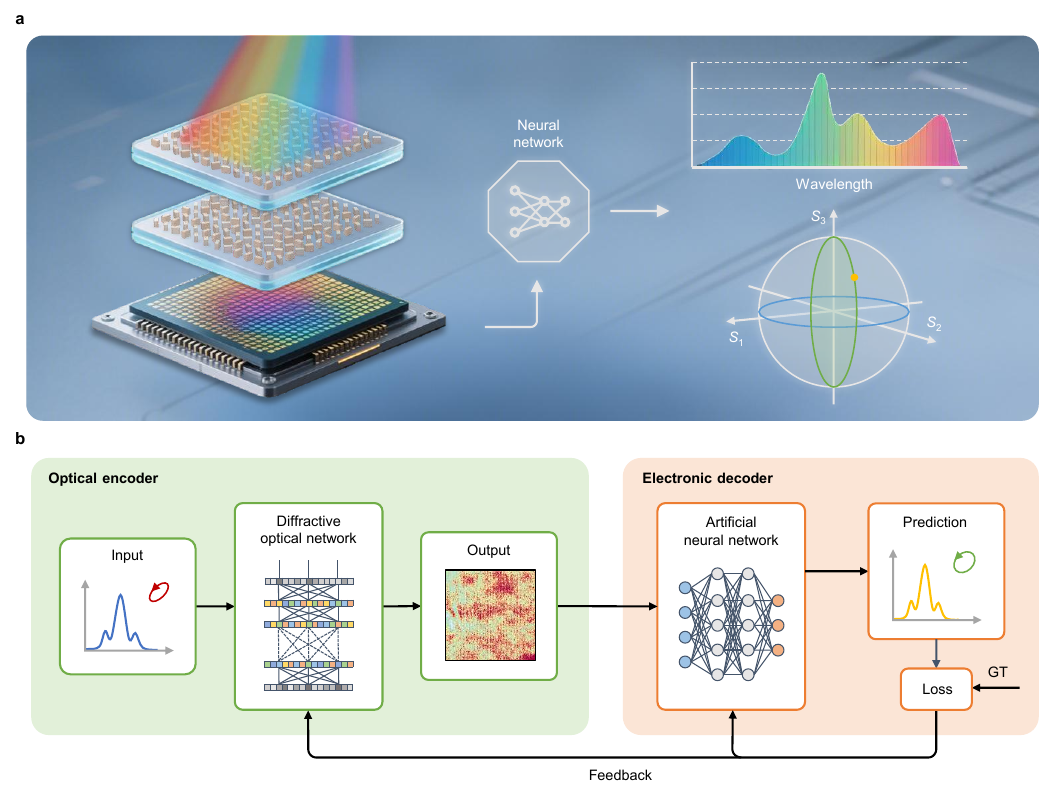}
\caption{\textbf{Integrated spectropolarimeter by metasurface-based DONs.} \textbf{a} The schematic of the DONs. Incident light with arbitrary spectral compositions and polarization states is encoded by the optical encoder, processed through the dual-layer cascaded metasurfaces, and captured as spatial intensity patterns on a CMOS sensor. These patterns are decoded by a co-designed electronic decoder to simultaneously reconstruct the spectrum and Stokes parameters. The metasurface-based DONs are ready for direct integration onto an image sensor chip. \textbf{b} The training workflow of the DONs. The physical parameters of the metasurfaces and weights of the artificial neural network are co-optimized by minimizing the loss function between predictions and ground truth (GT) through backpropagation.}\label{fig1}
\end{figure*}

DONs, also known as diffractive deep neural networks (D$^2$NNs), present a transformative solution by synergizing multi-layer metasurfaces with machine learning\cite{Lin2018}. Unlike conventional task-specific metasurface optimization, DONs employ end-to-end joint optimization of phase profiles and computational decoders. These innovative architectures employ multiple layers of diffractive elements that manipulate light to process high-dimensional information. 
Such DONs have demonstrated successful implementation in various intelligent tasks, such as image recognition and classification\cite{Yan2019, Liu2022, Luo2022}, generative model\cite{Zhan2024, Qiu2025}, information encryption\cite{Bai2024, Guo2025}, among many others\cite{Li2023, Mengu2023, Xu2024, Hu2024a, Qiu2024}. 

The schematic of the metasurface-based DONs for the integrated spectropolarimeter is illustrated in Fig. \ref{fig1}a. The DONs serve as a computational engine that transforms the encoded optical field into a machine-learnable representation, consisting of an optical encoder and an electronic decoder, enabling simultaneous extraction of spectral and polarization information. The optical encoder comprises single- or dual-layer cascaded metasurfaces, designed to transform the incoming high-dimensional spectral and polarization information into a spatially structured intensity distribution optimized for machine learning interpretation, forming a free-space optical processor that operates at the speed of light. By controlling the cumulative phase shifts across multiple diffraction planes, the DONs redistribute optical energy into distinct spatial channels that maximally separate different combinations of spectral and polarization states.
The electronic decoder is a convolutional neural network (CNN) simultaneously trained with metasurfaces that reconstructs the spectral information and Stokes parameters from the raw sensor intensity. The design of DONs integrates domain knowledge of optical physics to enhance robustness against noise and systematic errors.

Specifically, for an arbitrarily polarized light with spectral density $I(\lambda)$ is decomposed into left- and right-handed circularly polarized (LCP and RCP) components:
\begin{equation}
E_0(\lambda)= E_{l}(\lambda)\hat{e}_{l} + E_{r}(\lambda)\hat{e}_{r},
\end{equation}
where $\hat{e}_{l}=\frac{1}{\sqrt{2}}\binom{1}{i}$ and $\hat{e}_{r}=\frac{1}{\sqrt{2}}\binom{1}{-i}$. The metasurface imposes independent phase profiles on each circular polarization component, generating the encoded light field:
\begin{equation}
E(x,y,\lambda)= \sqrt{\eta(\lambda)}\left[E_{l}(\lambda)e^{i\phi_{l}(x,y,\lambda)}\hat{e}_l + E_{r}(\lambda)e^{i\phi_{r}(x,y,\lambda)}\hat{e}_r\right]+\sqrt{1-\eta(\lambda)}E_{0},
\end{equation}
where $\eta(\lambda)$ is the modulation efficiency at $\lambda$, $\phi_{l}$ and $\phi_{r}$ correspond to the phase shift for LCP and RCP, respectively. For the detailed derivation of polarization state decomposition, as shown in Supplementary Note 1. 

The encoded field propagates through the cascaded diffractive layers. The optical field $E_n$ at layer $n$ is calculated via angular spectrum propagation:
\begin{equation}
E_{n+1}(x,y) = \mathcal{F}^{-1}\left[ \mathcal{F}(E_n \cdot e^{i\phi_n(x,y)}) \cdot e^{i\frac{2\pi z}{\lambda}\sqrt{1-(\lambda f_x)^2 - (\lambda f_y)^2}} \right],
\end{equation}
where $z$ is the inter-layer spacing, $f_x$ and $f_y$ are spatial frequencies, and $\phi_n(x,y)$ is the phase profile of layer $n$.  

After passing through the last layer $N$, the intensity distribution $I(x,y)$ is captured by the CMOS detector:
\begin{equation}
I(x,y) = \int_{\lambda_{\min}}^{\lambda_{\max}} \left| E_N(x,y,\lambda) \right|^2 d\lambda,
\end{equation}
thus contains entangled spectral and polarization information, structured by the metasurfaces to facilitate neural network decoding.

Unlike conventional methods where optical components and algorithms are developed independently, our framework implements end-to-end differentiable optimization that co-designs nanophotonic structures and computational models, as shown in Fig. \ref{fig1}b. This is enabled by a custom computational graph that backpropagates gradients from the electronic decoder through the physical optical model. The phase profiles $\phi_n(x,y)$ of the diffractive layers are parameterized, while the decoder employs a CNN with residual blocks for robust feature extraction. The optimization process minimizes a hybrid loss function consisting of spectral fidelity and polarization accuracy. In addition, the optical forward model incorporates stochastic noise simulating CMOS readout noise and fabrication errors, ensuring the learned parameters are robust to real-world imperfections, see Methods for network details.

This co-design method offers a systematic framework for developing multi-functional photonic systems through bidirectional information flow between physical optics and machine learning. 
Our joint optimization allows bidirectional information exchange between physical layer responses and electronic network mappings.
During training, gradient signals from the decoder guide the phase profiles of metasurface updates toward configurations that maximize measurable feature discrimination, while the physical constraints of light propagation reciprocally regularize the model's weight space to avoid overfitting and thus shape the decoder's feature extraction strategy. This feedback loop enables concurrent enhancement of both optical encoding efficiency and electronic decoding robustness, the dual optimization optimization objective that remains inaccessible to traditional sequential design methods. The details of DONs' training process are shown in the supplementary Note 2. In addition, to further enable imaging, the dataset not only covers a wide range of spectral compositions and polarization states, but also incorporates spatially structured input patterns, including various irregularly shaped beams, during the training process to teach the network to associate local intensity variations with spectral and polarization features. This effectively transforms the DONs into a computational imaging device.

\subsection{Metasurface design and characterization}

\begin{figure*}[!ht]%
\centering
\includegraphics[width=\textwidth]{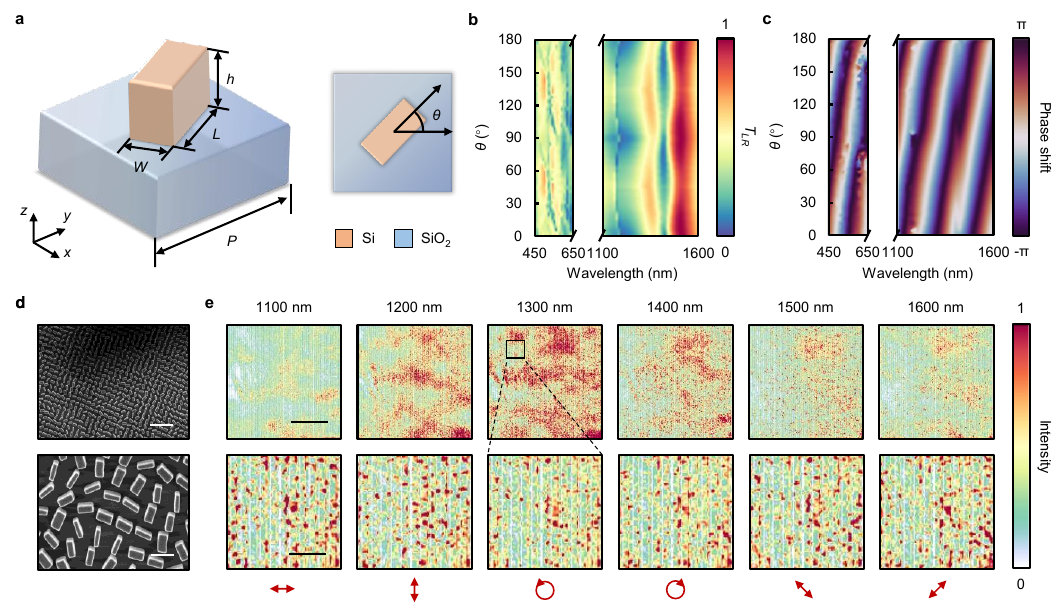}
\caption{\textbf{Design of metasurface-based encoder.} \textbf{a} The schematic of the Si nanobrick array. \textbf{b,c} The transmission and phase shift maps of the example nanobrick. \textbf{d} The large-area SEM image confirming uniform nanobrick distribution and high-resolution image of individual nanobricks. The scale bars are 2 $\upmu$m and 500 nm, respectively. \textbf{e} The experimentally measured intensity distributions under different wavelengths and incident polarization states, where the incident polarization state is LCP for different wavelengths (upper panel) and the wavelength is 1300 nm for different polarization states (lower panel). The scale bars are 100 $\upmu$m and 20 $\upmu$m, respectively.}\label{fig2}
\end{figure*}

The metasurface encoder serves as the core of DONs. To achieve polarization-dependent operation, we design anisotropic silicon (Si) nanobricks as meta-atoms, arranged in a square lattice $P$ with 600 nm periodicity on a silicon dioxide (SiO$_{2}$) substrate, as illustrated in Fig. \ref{fig2}a. Each nanobrick has a height $h$ of 700 nm, with width $W$ of 140, 200, 240, or 260 nm, length $L$ of 460 or 500 nm, and a rotation angle $\theta$. Each metasurface layer consists of a 500 $\times$ 500 array of meta-atoms, see Methods for sample fabrication details. 
We perform numerical simulations using the finite difference time domain (FDTD) solver to calculate the spectral response. As an example, Figs. \ref{fig2}b,c present the simulated transmission from LCP-to-RCP $T_{LR}$ and phase shift for meta-atoms with $W=200$ nm and $L=500$ nm. 
The results demonstrate broadband operation with full $0–2\pi$ phase coverage, satisfying the key requirements for DONs and these characteristics are maintained across other geometric parameters.

The phase modulation scheme combining propagation phase and geometric phase is adopted to decouple the wavefront control for LCP and RCP, which is a common approach for circular polarization multiplexing\cite{Li2020, Liu2022a}. Specifically, the variations in length and width of the anisotropic meta-atoms govern the propagation phase, while the rotation angle introduces a geometric phase for circularly polarized light. For circularly polarized incident light, the propagation phase of the output polarization-converted component is given by: 
\begin{equation}
\phi_p=\mathrm{arg}(\frac{t_x-t_y}{2}), 
\end{equation}
where $t_x=e^{(i\phi_{xx})}$ and $t_y=e^{(i\phi_{yy})}$, representing the phase delays along the $x$- and $y$-axes, respectively. Simultaneously, the geometric phase depends on the rotation angle $\theta$ and the helicity of the incident light, given by $\phi_g=\pm 2\theta$, the "+" and "–" signs corresponding to LCP-to-RCP and RCP-to-LCP conversions. The total phase for the polarization-converted component thus becomes:
\begin{equation}
\phi=\mathrm{arg}(\frac{t_x-t_y}{2}) \pm 2\theta.
\end{equation}

This design ensures distinct phase profiles for orthogonal circularly polarized incident beams. Since any arbitrary polarization state can be decomposed into these two orthogonal circularly polarized components, the metasurfaces generated spatial intensity patterns exhibit high sensitivity to incident polarization states. Fig. \ref{fig2}d shows scanning electron microscopy (SEM) images of the fabricated metasurfaces, demonstrating the precise fabrication of the nanostructures. Fig. \ref{fig2}e shows the representative intensity distributions captured by the camera under different wavelengths and incident polarization states, demonstrating the inherent ability of the metasurface to encode spectral and polarization information into distinct spatial intensity patterns. It can be seen that the patterns are significantly distinct at different wavelengths, while different polarization states at the same wavelength can still be distinguished, although the differences are relatively smaller.
These distinct intensity patterns from wavelength-dependent diffraction are governed by the polarization-sensitive phase modulation of the metasurface, which together ensure minimal crosstalk between the spectral and polarization channels. These spatially resolved optical fingerprints form the basis for subsequent neural network decoding of spectral and polarization state reconstruction.

\subsection{Experimental validation}

\begin{figure*}[!ht]%
\centering
\includegraphics[width=\textwidth]{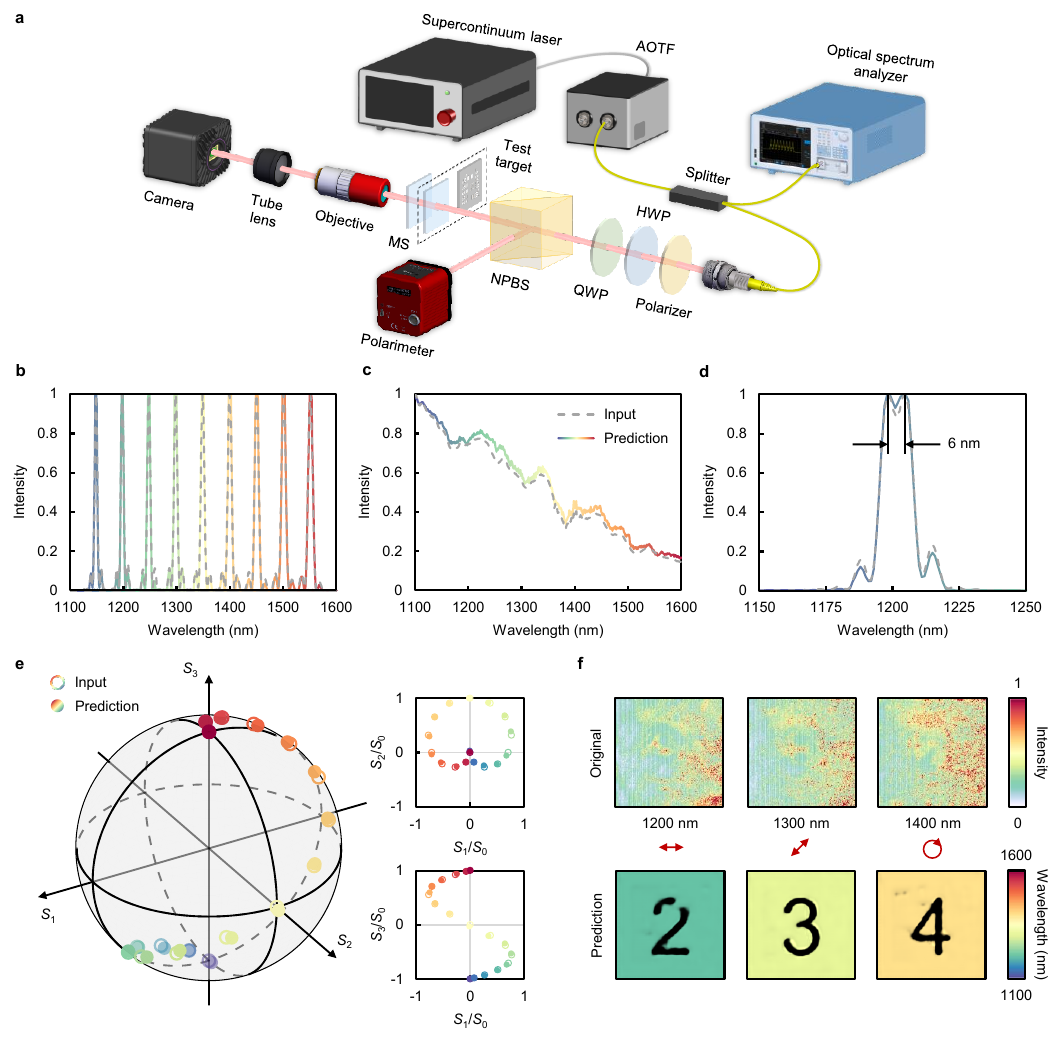}
\caption{\textbf{Experimental validation of spectropolarimeter.} 
\textbf{a} The schematic of the experimental setup. Samples are illuminated by wavelength- and polarization-modulated light, with spatial intensity patterns captured by a camera. Reference measurements are obtained using commercial optical spectrum analyzer and polarimeter. 
\textbf{b} The spectral reconstruction performance for discrete narrowband sources, demonstrating precise wavelength discrimination capability (MAE = 0.004).
\textbf{c} The broadband spectral reconstruction results showing excellent agreement between reconstructed and reference spectra, confirming polychromatic source compatibility  (MAE = 0.031, MRE = 7.39\%).
\textbf{d} The spectral resolution test using dual-peak measurements, demonstrating the ability to resolve spectral features separated by 6 nm (equipment-limited resolution).
\textbf{e} The full-Stokes polarimetric reconstruction performance evaluated using polarization states uniformly distributed across the Poincaré sphere, achieving an average MAE of 0.016 for Stokes parameters.
\textbf{f} Three digits encoded with distinct wavelength-polarization combinations are reconstructed via a neural network, yielding high-fidelity images with minimal background noise.
The scale bars are 100 $\upmu$m. 
HWP, half-wave plate; QWP, quarter-wave plate; NPBS, non-polarized beam splitter; MS, metasurface.}\label{fig3}
\end{figure*}

The spectropolarimeter by metasurface-based DONs are rigorously validated through a series of experiments that evaluating its spectrometric, polarimetric, and imaging capabilities in the near-infrared (NIR) band (1100--1600 nm).
The experimental setup is shown in Fig. \ref{fig3}a. The samples are illuminated by wavelength- and polarization-modulated light, and the intensity patterns are captured by a camera. Reference measurements are acquired by commercial polarimeter and optical spectrum analyzer, see Methods for experimental setup details.

To validate the sensing capabilities, we conduct comprehensive experiments using diverse spectral profiles and polarization states, demonstrating simultaneous reconstruction of both information.
Firstly, to verify the operation bandwidth of the spectropolarimeter, we use an acousto-optic tunable filter (AOTF) to generate a series of spectral lines in the range of 1100 nm to 1600 nm and reconstruct each one individually. As shown in \ref{fig3}b, all spectral lines have been successfully reconstructed with high precision, and the average mean absolute error (MAE) is 0.004.
For broadband light sources, the spectropolarimeter demonstrates remarkable fidelity as shown in Fig. \ref{fig3}c. The reconstructed spectrum exhibits almost overlap with the original spectrum, achieving an MAE of 0.031 and a mean relative error (MRE) of 7.39\% across the 500 nm bandwidth. This broadband consistency confirms the system's universal applicability to both narrowband and polychromatic light sources.
The resolution of the spectropolarimeter is evaluated by dual-peak measurement, where the DONs resolve narrowband spectral features with precision, distinguishing laser lines separated by 6 nm, as shown in Fig. \ref{fig3}d. Note that this is not the upper limit on the resolution, but rather the narrowest dual peaks that can be produced by our experimental equipment.

Fig. \ref{fig3}e presents experimental results demonstrating the reconstruction of the full Stokes parameters. The polarimetric performance of the spectropolarimeter is rigorously assessed by measuring polarization states uniformly distributed across the entire Poincaré sphere at various wavelengths within the operating spectral range. These measurements yield reconstructed Stokes parameters that achieve an average MAE of 0.016.

In addition, while the experimental results above validate the system’s ability to reconstruct spectrum under a fixed polarization state, numerical simulations further demonstrate that our DONs simultaneously resolve distinct polarization states carried by different wavelengths in polychromatic light, as detailed in the Supplementary Note 3.

To further validate the system's multi-dimensional imaging capabilities, we perform an imaging experiment employing three target digits, each digit with distinct combinations of wavelength and polarization states, as shown in Fig. \ref{fig3}f. This scenario can also be viewed as imaging through scattering medium, and we utilize an additionally neural network to recover the image, as detailed in the Supplementary Note 4. To reduce the difficulty of image reconstruction and sample calibration, we use a single-layer metasurface as the optical encoder, and see the Supplementary Fig. S5 for details on the performance comparison between single- and dual-layer. The reconstructed images are clear and have high fidelity with low background noise. The average mean square error (MSE) for image reconstruction is 0.004.

\subsection{Chip-integrated sensor implementation}

\begin{figure*}[!ht]%
\centering
\includegraphics[width=\textwidth]{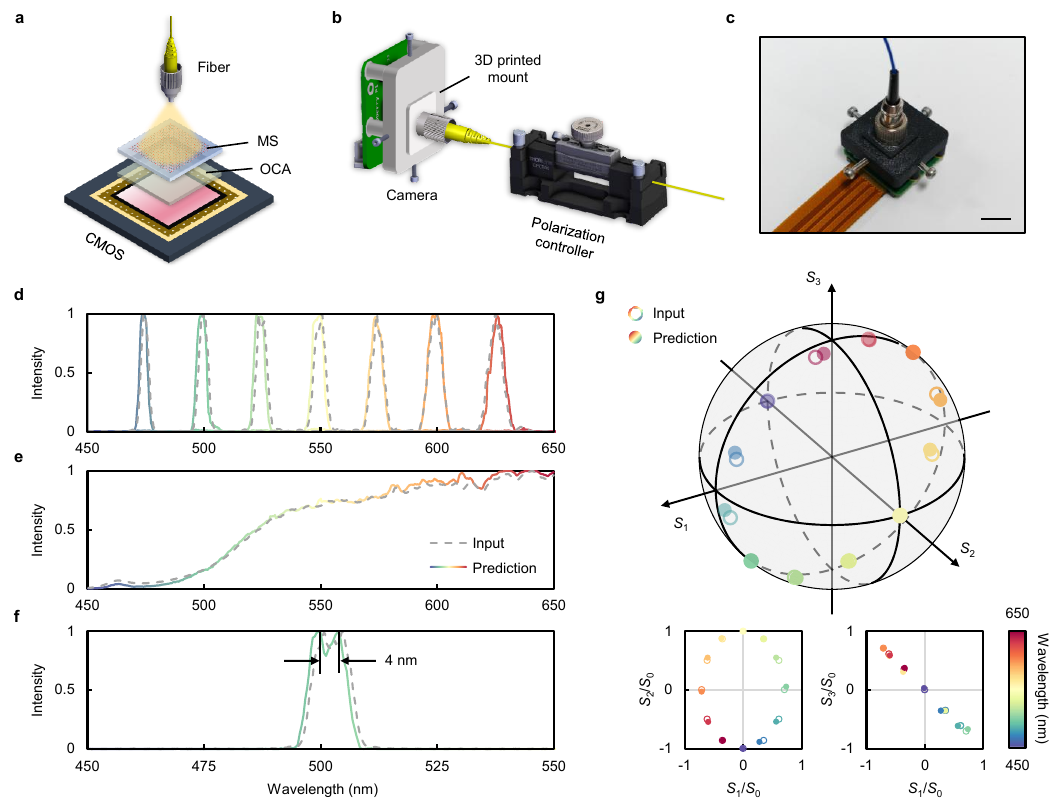}
\caption{\textbf{Chip-integrated metasurface-based DON sensor prototype.} 
\textbf{a} The architecture of chip-integrated sensor. The OCA is used to bond the metasurface encoder to a CMOS pixel array.
\textbf{b} The experimental setup with fiber-coupled source and 3D-printed alignment mount to simplify optical coupling. 
\textbf{c} The photo of the integrated sensor prototype, showing a pretty compact form factor. The scale bar is 1 cm.
\textbf{d--g} The performance characterization of the integrated sensor, demonstrating reconstruction fidelity for \textbf{d} narrowband (MAE = 0.007) and \textbf{e} broadband (MAE = 0.027, MRE = 6.98\%), with \textbf{f} 4 nm spectral resolution (equipment-limited), and \textbf{g} polarization reconstruction MAE = 0.039. The integrated sensor maintains competitive performance while enabling single-shot, compact spectropolarimetric sensing.}\label{fig4}
\end{figure*}

The transition from discrete optical components to a fully integrated sensor prototype is a crucial step to realize practical applications of metasurface-based devices, validated through systematic experiments using visible light sources.
The schematic of the sensor is shown in Fig. \ref{fig4}a, at the heart of this integration is a monolithic architecture that attaches the metasurface encoder directly to the pixel array of a commercial CMOS image sensor using optically clear adhesive (OCA), providing robust optical alignment while maintaining millimeter-scale compactness. The experimental setup is shown in Fig. \ref{fig4}b, where the modulated light is coupled directly to the metasurface encoder via a single mode fiber, and we create a three-dimensional (3D) printed mount to reduce alignment complexity. Fig. \ref{fig4}c shows a photograph of the sensor, which is almost the same size as a conventional CMOS imaging sensor. The experimental setup is detailed in Methods.

We further evaluate the chip-integrated sensor prototype under similar test conditions. As shown in Figs. \ref{fig4}d--g, the integrated sensor achieves a spectral resolution of 4 nm (equipment-limited resolution) and operates over a bandwidth of 450--650 nm, covering nearly the entire visible spectrum. It achieves an average MAE of 0.007 for narrowband source reconstruction, an MAE of 0.027 for broadband source reconstruction, and an MRE of 6.98\% in the range of 480 nm to 650 nm. The polarization state reconstruction has an average MAE of 0.039 over 12 polarization states.
Although direct comparison with previous experiments is precluded due to different operational wavelength ranges, the integrated sensor maintains competitive performance metrics, demonstrating successful translation from discrete optical devices to chip-scale implementation.

These experimental results collectively validate the proposed system's capability as a compact, single-shot, multi-dimensional optical sensor that achieves simultaneous high-fidelity spectropolarimetry while enabling new paradigms in intelligent photonic sensing.

\section{Discussion}\label{sec3}
Our proposed metasurface-based spectropolarimeter establishes a new paradigm for simultaneous spectrometric and polarimetric measurements. Compared to conventional systems that require bulky and discrete optical devices, our design achieves a chip-scale footprint while eliminating mechanical scanning or sequential measurements, which is critical for real-time applications such as dynamic sensing, while maintaining competitive spectral resolution and polarization accuracy. 
This performance is enabled by the wavelength- and polarization-dependent phase modulation capability of the metasurface encoder, which projects high-dimensional optical information onto an image sensor plane. In addition, the chip-integrated prototype further demonstrates the feasibility of this approach in practical applications.
The key innovation lies in the co-design of physics-driven metasurfaces and data-driven deep learning, which synergistically encode and decode multi-dimensional optical information. 
Unlike conventional metasurfaces designed for specific functionalities, our proposed DONs inherently resolves the coupling between spectral and polarization responses through joint design of metasurface phase profiles and weights of artificial neural network.

Looking ahead, the seamless fusion of metasurface optics with semiconductor sensors opens transformative opportunities. Future iterations may incorporate reconfigurable metasurfaces to dynamically adjust phase profiles, or co-optimize optical encoding and electronic readout at the hardware level, suppressing noise while enhancing dynamic range\cite{Fan2024}. 
In addition, beyond spectropolarimetric sensing and imaging, such platforms could unify beam quality analysis\cite{Zhou2021, Gu2024}, orbital angular momentum\cite{Guo2021, Zhang2023}, and other optical modalities into all-in-one optical sensor processors. 
This integrated metasurface-based DON framework bridges nanophotonics and computational sensing,  with great potential for a wide range of applications.

\section{Methods}\label{sec4}
\subsection{Network details}
We use the open-source machine learning framework Pytorch to build the algorithm. The training process is implemented using Python (v3.10.10) and Pytorch (v2.0.0) framework on a desktop computer with Intel Core i7-13700K CPU, Nvidia GeForce RTX 4070 Ti GPU, and 32 GB RAM.

We train each network independently for its respective task, maintaining an identical architecture across all networks. The optical encoder comprises either a single- or dual-layer metasurface design. Within each metasurface layer, a dense array of $500 \times 500$ subwavelength nanostructures (functionally analogous to artificial neurons) modulates the incident light. These layers are separated by a propagation distance of approximately 1 mm within the optical path.

The electronic decoder is primarily composed of convolutional layers, including a series of convolutional layers with 1$\times$1 and 3$\times$3 kernels, batch normalization, and leaky rectified linear unit (LeakyReLU) activation functions. 
The network employs residual connections to mitigate vanishing gradients and enable deeper training. These residual blocks typically stack multiple convolutions, allowing the model to learn more complex features efficiently.
The network progressively downsamples the input through strided convolutions, increasing channel depth while reducing spatial dimensions. 
We resize the intensity patterns to 500$\times$500 and input to the network. The network finally generates a 16$\times$16$\times$1024 shaped feature map, followed by global pooling for each channel of the feature map and fully connected layers to generate the spectral and polarization information. 
The schematic of the network is shown in Supplementary Fig. S1.



\subsection{Sample fabrication}
The metasurface fabrication process begins with the deposition of a 700 nm thick Si layer on the SiO$_{2}$ substrate via ion beam-assisted deposition (IBAD), followed by a 25 nm thick chromium (Cr) hard mask is deposited using electron beam thermal evaporation. Then, a 300 nm thick electron beam resist layer is spin-coated onto the hard mask layer. After soft baking, electron-beam lithography patterns the resist, which is developed to form protective pillar arrays. This resist pattern is transferred into the underlying chromium via reactive ion etching, creating a durable hard mask. Following resist stripping, the Si layer is etched into high-aspect-ratio nanopillars using inductively coupled plasma reactive ion etching (ICP-RIE), and the Cr mask is used as an etch barrier. Finally, wet etching removes the Cr mask, leaving precisely defined Si nanopillars on the transparent SiO$_{2}$ substrate. The schematic of the process is shown in Supplementary Fig. S6.

\subsection{Experimental setup}
In the NIR experimental demonstration, the optical system uses a supercontinuum laser (SC-Pro, YSL Photonics) as the broadband light source. Spectral tuning across the visible to NIR range is achieved using an AOTF (AOTF-Pro2, YSL Photonics), which enables simultaneous output of eight discrete wavelengths with a full width at half maximum (FWHM) of 2--10 nm. The filtered light is subsequently divided by a fiber splitter, with one output path dedicated to monitoring the spectral characteristics using an optical spectrum analyzer (AQ6370D, Yokogawa). The other path passes through a polarization conditioning system, where the light passes sequentially through a polarizer, a HWP, and a QWP to generate arbitrary polarization states. The polarized beam is then directed through a 50:50 split ratio NPBS, creating measurement and sampling paths. The measurement path terminates at a polarimeter (PAX1000IR2/M, Thorlabs) for polarization state analysis, while the sampling path first interacts with a test target before illuminating the samples. Finally, the scattered light from the metasurface samples is collected by an objective (M Plan Apo NIR 20×, Mitutoyo) and imaged through a tube lens system onto a shortwave infrared (SWIR) camera (C-RED 2 Lite, First Light). The metasurface samples are mounted on three-axis motorized stages with piezo linear actuators (8301NF, New Focus) to ensure sub-nanometer alignment accuracy. A self-developed Python script controls the above devices and collects data.

The implementation of the chip-integrated sensor retains the core components described above. The difference is that we use a fiber optic polarization controller (CPC900, Thorlabs) to adjust the output polarization state, a fiber optic spectrometer (USB4000, Ocean Optics) to measure the spectrum, and the rotating QWP method to measure the polarization state. The image sensor is a Sony IMX219 with 1.12 $\upmu$m pixel pitch in a 1/4 inch optical format, which provides sufficient resolution for the detection of spatial intensity patterns. Although primarily designed for RGB imaging, its monochromatic imaging capability meets resolution requirements while offering significant cost advantages. The optical fiber is attached to the sensor PCB via a 3D printed polylactic acid (PLA) mount, and the OCA layer thickness is 250 $\upmu$m. The fiber's ceramic ferrule is secured by a plastic clamp with precision alignment screws, similar to the design of FiberPort, providing micrometer-level alignment stability between the fiber core and the metasurface centroid.

\section*{Data availability}

All data are available in the main text and the Supplementary Information.



\section*{Acknowledgments}

This work was supported by the National Natural Science Foundation of China (Nos. 12064025, 12264028, 12364045, 12304420, and U24A20304), the Natural Science Foundation of Jiangxi Province (Nos. 20212ACB202006, 20232BAB201040, and 20232BAB211025), Young Elite Scientists Sponsorship Program by JXAST (Nos. 2023QT11 and 2025QT04), and Jiangxi Provincial Key Laboratory of Photodetectors (No. 2024SSY03041).

\section*{Author Contributions}

J.Q. and T.L. conceived the idea. J.Q., and C.X. performed the theoretical calculation and numerical simulations. J.Q. and S.X. performed the experiments. J.Q., T.L., and T.Y. prepared the manuscript. T.Y., Q.L., and S.X. supervised the project. All authors discussed the results and approved the manuscript.

\section*{Competing Interests}

The authors declare no competing interests.

\end{document}